 \documentclass[pmlr,10pt]{jmlr} 

\usepackage{booktabs}
\usepackage{siunitx}

\usepackage[switch]{lineno}

\theorembodyfont{\upshape}
\theoremheaderfont{\scshape}
\theorempostheader{:}
\theoremsep{\newline}

\jmlrvolume{XXX}
\jmlryear{2026}
\jmlrworkshop{} 

 \title[RamanSeg]{RamanSeg: Interpretability-driven Deep Learning on Raman Spectra for Cancer Diagnosis}

\author{%
\Name{Chris Tomy\nametag{\thanks{ct678@cantab.ac.uk}}}\\
\addr Department of Computer Science and Technology, University of Cambridge, Cambridge, UK\\\\
\Name{Mo Vali}\\
\addr Department of Physics, University of Cambridge, Cambridge, UK\\\\
\Name{David Pertzborn}, \Name{Tammam Alamatouri}, \Name{Anna M\"{u}hlig}, \Name{Orlando Guntinas-Lichius}\\
\addr Department of Otorhinolaryngology, Jena University Hospital, Jena, Germany\\\\
\Name{Anna Xylander}\\
\addr Section Pathology of the Institute of Forensic Medicine, Jena University Hospital, Jena, Germany\\\\
\Name{Eric Michele Fantuzzi}, \Name{Matteo Negro}, \Name{Francesco Crisafi}\\
\addr Cambridge Raman Imaging Srl, Milan, Italy\\\\
\Name{Pietro Lio}, \Name{Tiago Azevedo\nametag{\thanks{tiago.azevedo@cst.cam.ac.uk}}}\\
\addr Department of Computer Science and Technology, University of Cambridge, Cambridge, UK
}

\begin{document}

\maketitle

\begin{abstract}
Histopathology, the current gold standard for cancer diagnosis, involves the manual examination of tissue samples after chemical staining, a time-consuming process requiring expert analysis.
Raman spectroscopy is an alternative, stain-free method of extracting information from samples.
Using nnU-Net, we trained a segmentation model on a novel dataset of spatial Raman spectra aligned with tumour annotations, achieving a mean foreground Dice score of 80.9\%, surpassing previous work.
Furthermore, we propose a novel, interpretable, prototype-based architecture called \textit{RamanSeg}.
RamanSeg classifies pixels based on discovered regions of the training set, generating a segmentation mask.
Two variants of RamanSeg allow a trade-off between interpretability and performance: one with prototype \textit{projection} and another \textit{projection-free} version.
The projection-free RamanSeg outperformed a U-Net baseline with a mean foreground Dice score of 67.3\%, offering a meaningful improvement over a black-box training approach.
\end{abstract}




\section{Introduction}
\label{sec:intro}

To combat cancer, diagnosis is a critical first step in treatment.
Histopathology, the examination of tissue samples, is currently the gold standard for diagnosis~\citep{aljehani7importance}.
To assist inspection, a chemical stain is applied, for example, the \textit{hematoxylin and eosin} (H\&E) stain~\citep{chan2014hematoxylin}.
This is time-consuming. As a stain-free alternative, \textit{Raman spectroscopy} directly provides molecular information by measuring scattered laser light from a tissue sample.
While this spectral data is difficult to use directly, we could train deep learning models via a supervised training setup.
Hence, this work utilizes a novel dataset from the EU-funded CHARM project~\citep{charm_eic}, composed of spatial Raman data along with tumour annotations.
Previous work~\citep{Hollon2020} investigated the performance of a dataset composed of two peaks of the Raman spectrum; for this work, we train over a newer dataset with 21 peaks spread over the \textit{C-H stretching region} of the Raman spectrum.

We adapted two existing convolutional neural network (CNN) architectures: U-Net~\citep{ronneberger2015unetconvolutionalnetworksbiomedical} and ProtoPNet~\citep{chen2018}.
Firstly, we applied the \textit{nnU-Net} framework to establish a strong baseline, improving on initial results with hand-written U-Net and UNet Transformer models~\citep{hatamizadeh2022unetr}. Our nnU-Net model achieved a mean foreground Dice score of 80.9\%, in contrast to a score of 72\% on the previous two-peak dataset.
Secondly, we developed a novel approach called \textit{RamanSeg}. RamanSeg is an adaptation of the ProtoSeg architecture~\citep{sacha2023protoseg}, which is itself an adaptation of the ProtoPNet architecture~\citep{chen2018}. Notably, we alter the objective function, introducing an \textit{activation overlap loss}, making it more computationally efficient to train. Additionally, our RamanSeg approach has two variants: one that includes a step called \textit{prototype projection}, and one that does not, representing a meaningful difference in performance and interpretability.

Overall, we make the following contributions:
\begin{itemize}
    \item The first successful application of a segmentation model to spatial Raman data across the entire C-H stretching region. Our model achieved a mean foreground Dice score of 80.9\%.    
    \item An empirical demonstration that architectures with a latent bottleneck (like prototype-based architectures) have the capability to generate high-quality segmentation masks.
    \item \textit{RamanSeg}, an adaptation of the ProtoSeg architecture, with an efficient, novel objective function and training process, exploring the potential for interpretable segmentation models on spectral data collected for cancer diagnosis.
\end{itemize}








\section{Related Work}
\label{sec:related-work}

Early work by \citet{mahadevan1996raman} investigated the application of Raman spectroscopy for tissue samples. Specifically, they demonstrated that the intensities of peaks in the Raman spectrum can be used to distinguish between normal, precancerous, and cancerous tissue samples.
Their research focused on the fingerprint region of the Raman spectrum, approximately between 300 to 1900 cm$^{-1}$.

More recent work has investigated using the stretching region, approximately 2800 to 3100 cm$^{-1}$.
For example, \citet{hed.24477} used the stretching region as part of their analysis of head and neck squamous cell carcinoma.
\citet{Hollon2020} looked at two notable peaks within the stretching region, 2850 and 2950 cm$^{-1}$: peaks known to indicate the presence of lipids and cellular regions respectively.
Previous work internal to the CHARM group involved training a segmentation model using a dataset of the same two peaks, over head and neck tissue samples.
This model achieved a Dice score of 72\%, and represents a useful comparison point to judge our work, which instead leverages the full range of the stretching region.

The ability to interpret models is often a requirement for clinical deployment.
Traditionally, \textit{post-hoc} interpretability techniques have been used to understand CNNs. 
Some techniques include saliency maps~\citep{rudin2018} or Gradient-weighted Class Activation Mappings~\citep{selvaraju2020grad}.
A new paradigm of \textit{prototype-based} model architectures has emerged with \citet{chen2018}'s work.
The \textit{Prototypical Part Network} (ProtoPNet) classifies inputs based on \textit{prototypes} learned from the training set. Given a trained ProtoPNet, the model takes an input $\mathbf{x}$, maps it to a convolutional output $f(\mathbf{x}) = \mathbf{z}$, and computes distances $||\mathbf{\tilde{z}} - \mathbf{p}_j||$ between latent patches $\mathbf{\tilde{z}}$ and prototypes $\mathbf{p}_j$.
These distances are inverted to represent similarity and are used to generate the final classification probabilities.
\citet{sacha2023protoseg} extend this idea to an architecture that generates segmentation masks, called \textit{ProtoSeg}.

Our work brings together ideas from early research into spectroscopic-based techniques for cancer diagnosis, as well as this recent work making deep convolutional models interpretable.

\section{Method}
\label{sec:method}

\subsection{Experimental setup}
\label{sec:experimental-setup}

\paragraph{Hyperspectral dataset}
Tissue samples of 10 patients with Squamous Cell Carcinoma (SCC) were collected and preserved as Formalin-Fixed Paraffin-Embedded tissue (FFPE). Each FFPE sample is then sliced into thin sections, which are mounted on glass slides for Raman imaging and H\&E staining; in total, we used 32 samples. In the context of the EU-funded CHARM project~\citep{charm_eic}, we used a stimulated Raman microscopy (SRS) system~\citep{Crisafi2023} designed to overcome the challenges of existing commercial SRS systems, which often limit detection to one or two frequencies at a time. This microscope features an all-fiber, dual-wavelength, self-synchronized laser with a detection unit based on a multi-channel amplifier. This setup enables broadband SRS in the C-H stretching region, allowing for the simultaneous acquisition of up to 38 channels in 400ns.

For each sample, the microscope hardware collected 21 channels of Raman data in the C-H stretching region, with each channel corresponding to a different wavenumber within the range 2802 to 3094 cm$^{-1}$. Hence, we effectively have a vector of 21 Raman intensities for each pixel in the image, forming our main training input: $(N, C, H, W)$.
Additionally, the full dataset includes 24 channels made up of: 21 of the captured Raman peaks, a transmission channel representing light intensities, a \textit{Two-Photon Excitation Fluorescence} (TPEF) channel, and a \textit{Second Harmonic Generation} (SHG) channel.
TPEF is based on the absorption of two photons by a molecule, which excites it to a higher energy state~\citep{so2000two}.
SHG relies on the effect where two photons of the same frequency combine to produce a photon of double the frequency, useful for probing collagen structures in human tissue~\citep{han2005second}.
Normalization was performed by scaling values between the 5th and 95th percentiles into a $[0, 1]$ range for Raman channels, and between 1st and 99th percentiles for TPEF and SHG channels. Values outside the percentile range were mapped to zero or one as appropriate. Furthermore, we run k-means clustering on the H\&E image to identify foreground pixels. Then, the mean spectral profile of foreground pixels is applied column-wise to correct for intensity drifts during optical collection.

Pathologist annotations were provided in QuPath~\citep{bankhead2017qupath}, which we exported.
These included multiple classes, but we chose to train over two classes, with tumorous or necrotic regions mapped to the foreground class and all other tissue classes designated as background.
This was because of an imbalance of classes present in the annotations, making the binary setup more effective than k-class training.

In summary, the data we used includes three key components: (1) the multi-channel, spatial Raman spectra of tissue samples, collected using Raman spectroscopy, (2) the H\&E stained images of the corresponding samples, and (3) the tumour annotations (from pathologists) which segment the input. \figureref{fig:e341-sample} shows these three sources for a sample from the dataset.

\begin{figure*}[htbp]
\floatconts
  {fig:e341-sample}
  {\caption{From left-to-right: (a) Raman intensities of approximately 2910 cm$^{-1}$ for a single sample, (b) the H\&E stained image of the same sample, and (c) the H\&E stained image overlaid with tumor annotations from pathologists.}}
  {%
    \subfigure[Raman channel]{\label{fig:raman-7}%
      \includegraphics[width=0.3\linewidth]{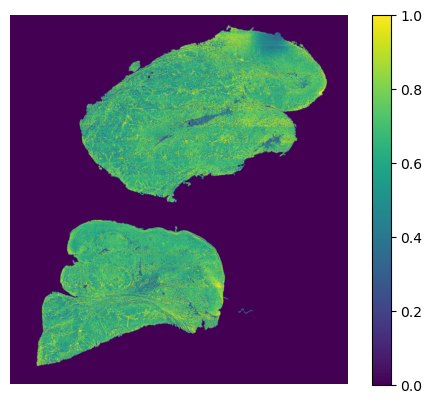}}%
    \qquad
    \subfigure[H\&E staining]{\label{fig:hne-e341}%
      \includegraphics[width=0.25\linewidth]{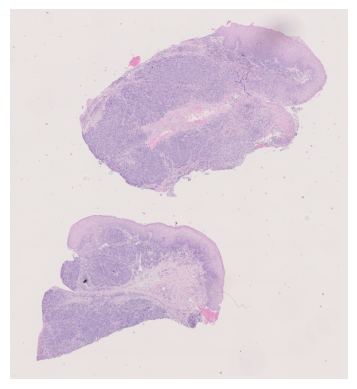}}
    \qquad
    \subfigure[Annotated staining]{\label{fig:hne-annotated}%
      \includegraphics[width=0.25\linewidth]{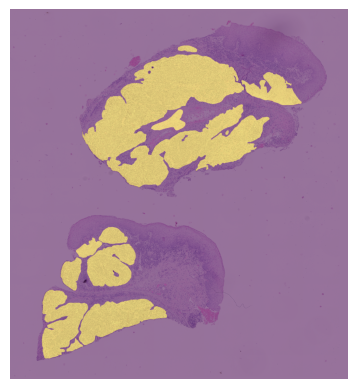}}
  }
\end{figure*}

\paragraph{Training} 
Models were either written in PyTorch, or using the nnU-Net library~\citep{isensee2021nnu}.
All training occurred on an NVIDIA L4 GPU, and for our final train, validation, and test
splits we used 19 training samples, 8 validation samples, and 5 test samples. To avoid data leakage, we grouped together samples from the same patient to be in the same split.
The U-Net~\citep{ronneberger2015unetconvolutionalnetworksbiomedical} architecture was implemented in PyTorch as a baseline.
This featured four downsampling layers, and we trained with a $512 \times 512$ patch size.
Our U-Net was trained with Dice loss, using the AdamW optimizer~\citep{loshchilov2017decoupled}, a learning rate of \verb|0.0001| and weight decay of \verb|0.001|. We used data augmentation: random horizontal flips of patches.

We also utilized the nnU-Net framework, selecting a Residual Encoder-based U-Net architecture. It featured seven downsampling stages and the framework chose a $576 \times 448$ patch size.
nnU-Net used a combination of Dice loss and cross-entropy loss, with an SGD optimizer, Nesterov momentum set to 0.99, and a learning rate of \verb|0.0001|.
Notably, we used nnU-Net to train five folds, with segmentation masks generated as an ensemble of the models.


\subsection{RamanSeg}

\begin{figure*}[htbp]
\floatconts
  {fig:ramanseg-arch}
  {\caption{The architecture of RamanSeg. The model consists of convolutional layers, producing latent feature maps which are passed to the prototype layer to generate similarity before probability maps.
    Finally, the probability map is upsampled to the original input size.
    (Diagram adapted from ProtoPNet~\citep{chen2018}.)}}
  {\includegraphics[width=\linewidth]{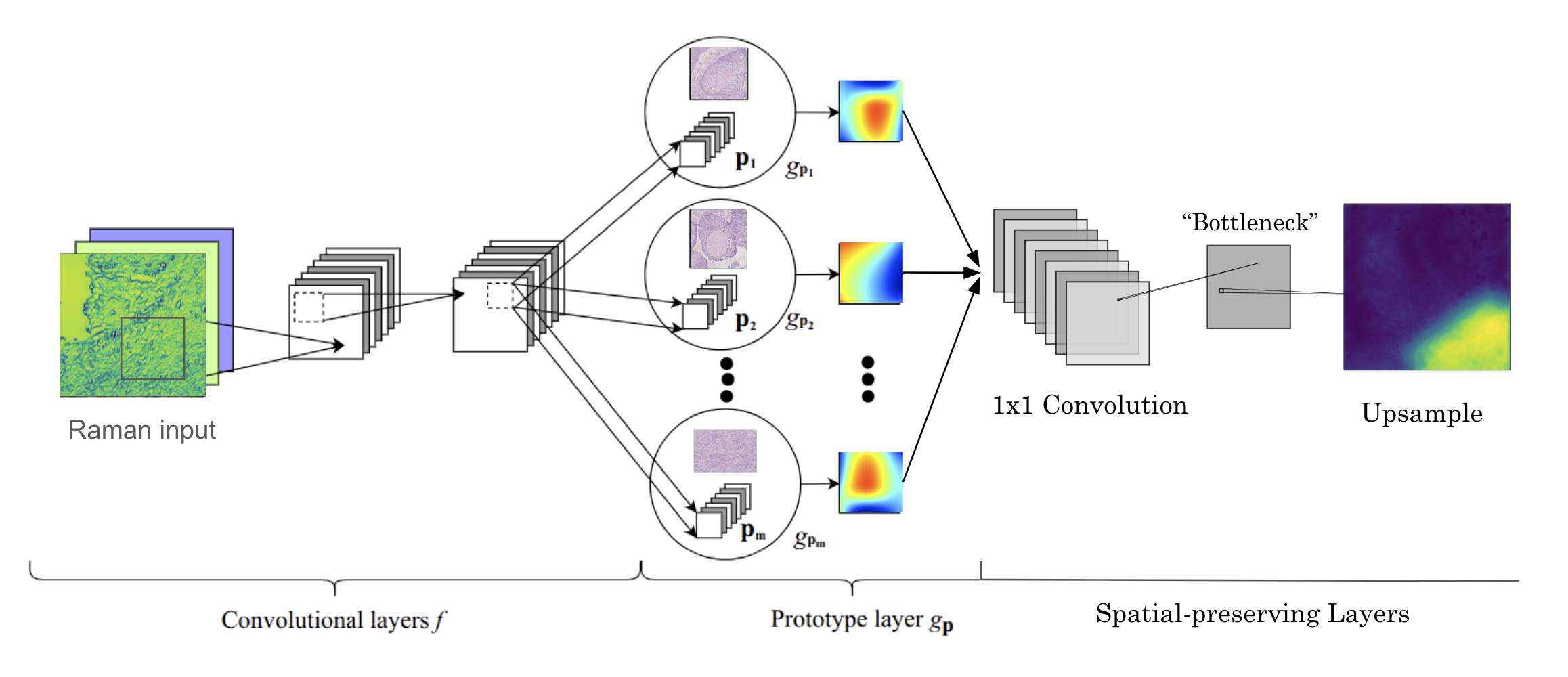}}
\end{figure*}

While the original ProtoSeg architecture by \citet{sacha2023protoseg} is an effective approach to prototype-based segmentation, we introduced a number of changes to the architecture and training process, leading to the development of RamanSeg, visualized in \figureref{fig:ramanseg-arch}.

\subsubsection{Bottleneck hypothesis}
\label{sec:bottleneck-hypothesis}

Similar to ProtoSeg~\citep{sacha2023protoseg}, RamanSeg takes the latent prototype similarity map of shape $H_d \times W_d \times M$ (for $M$ prototypes), passes each point through a fully connected layer, and outputs $H_d \times W_d \times C$ probabilities for each class.
Finally, it takes this map of per-point class probabilities and upscales it to the original input size $H \times W$, simply through bilinear interpolation.
This means it does not have a decoder: it relies on the ``quality" of the latent probability map to produce a segmentation mask.

Tumour segmentation masks can be fine-grained, so to verify if this architecture is viable for our dataset, we devised an experiment: assuming the best possible bottleneck, compute the Dice score between that upsampled bottleneck and ground truth mask.
Specifically, we downsampled the ground truth mask to a smaller size (representing the perfect $H_d \times W_d$ bottleneck), before upsampling it to the original size $H \times W$ with bilinear interpolation.
The results are in \tableref{fig:proto-bottleneck}, showing the average dice score for different bottleneck sizes.
While the bottleneck size does impact performance, we generally see high Dice scores, demonstrating that the prototype-based architecture is not fundamentally flawed on our dataset.

\begin{table}[hbtp]
\centering
\caption{Average dice scores for different bottleneck sizes. The original patch size is $512 \times 512$.}
\label{fig:proto-bottleneck}
\begin{tabular}{cc}
\toprule
\bfseries Feature Map Size & \bfseries Mean Dice \\
\midrule
$512 \times 512$ (Baseline) & 1.0000 \\
$32 \times 32$              & 0.9804 \\
$16 \times 16$              & 0.9581 \\
$8 \times 8$                & 0.9113 \\
$4 \times 4$                & 0.8356 \\
\bottomrule
\end{tabular}
\end{table}

\subsubsection{Activation overlap loss}
\label{sec:activation-overlap}

ProtoSeg~\citep{sacha2023protoseg} includes a penalty term in the objective function that incentivizes prototypes to be diverse, but their formulation of this diversity loss is computationally expensive.
The authors use the Kullback-Leibler divergence to measure the difference between distributions of prototype activations, but we instead formulate an alternative penalty term, called the \textit{activation overlap loss}.

Firstly, the forward pass through the model computes a similarity map $S$ of shape $(H_d, W_d, M)$, where $M$ is the number of prototypes; let $S'$ be the flattened similarity map, i.e. $(H_d \cdot W_d, M)$.
Each class has some associated prototypes (in our binary case, foreground and background), let the similarity map for prototypes of class $c$ be denoted as $S'_c$.
The key part in computing our activation overlap loss is the pairwise dot product between the similarity vectors (after flattening) for each class; the activation overlap loss term $L_A$ is shown in \equationref{eq:activation-overlap}.
\begin{equation}\label{eq:activation-overlap}
    L_A^{(c)} = \frac{\sum_{i \neq j}{S_c^{'(i)} \cdot S_c^{'(j)}}}{\binom{K}{2}} \quad{} L_A = \frac{1}{C}\sum_c{L_A^{(c)}}
\end{equation}

This makes computation fast as pairwise dot products are simpler than pairwise KL-divergence on ``prototype-class-image distance vectors''~\citep{sacha2023protoseg}. This leads to RamanSeg's overall objective function in \equationref{eq:ramanseg-objective}: a combination of cross-entropy, the activation overlap penalty, and an L1 penalty.
\begin{equation}\label{eq:ramanseg-objective}
    L = \alpha L_{\text{CE}} + \beta L_A + \gamma L_{\text{L1}}
\end{equation}

\subsubsection{Model \& training choices}
\label{sec:training-adaptations}

To enrich the model with additional capacity, we include a series of two-dimensional convolutions after the prototype layer, with Spatial Dropout~\citep{tompson2015efficient} at dropout probability $p = 0.35$ between them.
To improve training stability, we adapt the Xavier initialization scheme~\citep{glorot2010understanding} to prototype vectors.
For a prototype vector of shape $(D, H_P, W_P)$, values are drawn from the uniform distribution defined in \equationref{eq:xavier-proto}.

\begin{equation}\label{eq:xavier-proto}
    b = \frac{1}{\sqrt{D \cdot H_P \cdot W_P}} \quad \mathbf{p} \sim{} U(-b, b)^{D \times H_P \times W_P}
\end{equation}

Selecting hyperparameters is also particularly important for this architecture, for example, the number of prototypes $M$ to use.
Prior work on prototype-based architectures suggest using 10 prototypes per class~\citep{chen2018, sacha2023protoseg}, but we hypothesized more prototypes could be effective, given the per-pixel binary classification setup.
Fixing one variant of the RamanSeg architecture, we performed a hyperparameter search over the number of prototypes, within the range $[20, 200]$. \figureref{fig:proto-acc-vs-count} demonstrates that increasing the number of prototypes per class can offer better performance on our dataset.
Leveraging Optuna~\citep{akiba2019optuna} for a wider hyperparameter search, our final run in \sectionref{sec:results} used 15 prototypes per class.

\begin{figure}[h!]
\floatconts
  {fig:proto-acc-vs-count}
  {\caption{Performance when only varying the number of prototypes.}}
  {\includegraphics[width=0.7\linewidth]{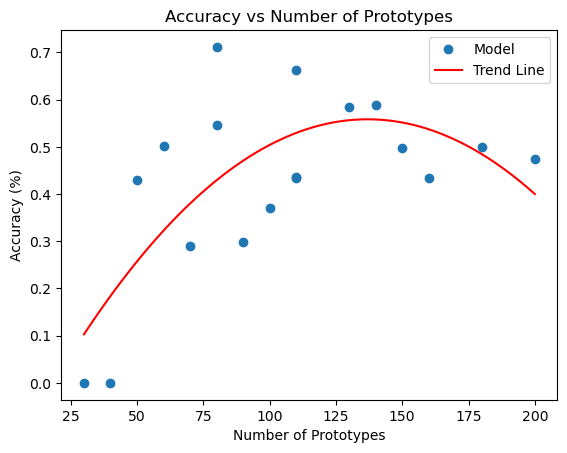}}
\end{figure}

\subsubsection{RamanSeg without projection}

As an alternative training approach, we propose \textit{projection-free RamanSeg}, which foregoes the prototype projection step discussed in previous prototype-based architectures~\citep{chen2018, sacha2023protoseg}.
We make further changes to the architecture under this projection-free variant: increasing the prototypes per class to 60, reducing the number of channels in the add-on convolutional layers (to 128), and increasing the dropout probability to 0.5.
Furthermore, there are two changes we consider particularly important (1) using a combined cross-entropy loss and Dice loss objective function, and (2) increasing the spatial size of the prototypes from $1 \times 1$ to $3 \times 3$. We elaborate on these choices below.

The standard RamanSeg training computes losses between the \textit{downsampled} ground truth mask and latent probability map.
(Note that the latent probability map, the ``bottleneck", is directly upsampled to generate the segmentation mask --- there is no learnable decoder.)
Hence, cross-entropy loss is critical to ensure the per-pixel correctness in the latent map. However, we found empirically that the combined loss performs better in training projection-free RamanSeg.
Secondly, we increased the spatial size of the prototypes to represent more complex features of the input space.
Hence, prototypes in projection-free RamanSeg have $(64, 3, 3)$ --- no longer 64-dimensional vectors.

By not projecting prototypes to their nearest representative region from the training set, we allow such richer representations. However, this makes prototypes more abstract, weakening the usual interpretability guarantees from inspecting similarity scores (which would otherwise correspond to similarity for specific prototypical regions).

\section{Results}
\label{sec:results}

The mean and standard deviation Dice scores of the trained models on the holdout dataset is presented in \tableref{fig:results}.
Given the medical context, we also compute the per-pixel sensitivity and specificity of the models.
The nnU-Net model outperforms other models, achieving a mean Dice score of 80.9\%.
Even with our challenging dataset, \citet{isensee2024nnu}'s claim that their approach is a ``recipe" for strong performance, appears to hold.
However, the nnU-Net model faces some interpretability challenges, outlined in \sectionref{sec:interp-analysis}.

On the other hand, projection-free RamanSeg achieves a mean Dice score of 67.3\%, outperforming the baseline U-Net model.
While omitting the projection step somewhat reduces interpretability, the bottleneck approach still provides greater interpretability than traditional models like U-Net, delivering a practical architecture that balances interpretability and performance.

\begin{table*}[hbtp]
\floatconts
  {fig:results}
  {\caption{Results of our trained models on the holdout set.}}
  {
    \begin{tabular}{lccc}
    \toprule
    \bfseries Model & \bfseries Dice ($\pm$ std) & \bfseries Sensitivity ($\pm$ std) & \bfseries Specificity ($\pm$ std) \\
    \midrule
    U-Net & 66.7 $\pm$ 15.4 & 81.5 $\pm$ 23.6 & 90.9 $\pm$ 5.5 \\
    UNet Transformer & 69.8 $\pm$ 10.9 & 79.9 $\pm$ 18.1 & 92.5 $\pm$ 3.9 \\
    nnU-Net & \textbf{80.9 $\pm$ 10.4} & \textbf{83.5 $\pm$ 14.5} & \textbf{95.9 $\pm$ 2.8} \\
    RamanSeg & 60.5 $\pm$ 11.7 & 95.9 $\pm$ 3.4 & 79.1 $\pm$ 4.9 \\
    Projection-free RamanSeg & 67.3 $\pm$ 8.2 & 70.3 $\pm$ 17.5 & 93.3 $\pm$ 5.5 \\
    \bottomrule
    \end{tabular}
  }
\end{table*}

\subsection{Interpretability analysis}
\label{sec:interp-analysis}

In the following sections, we investigate the differences in interpretability between the two model paradigms: the traditional CNN model via nnU-Net, compared to our prototype-based architecture, RamanSeg.

\subsubsection{nnU-Net interpretability analysis}

One of the behaviours we identified was the presence of consistent false positives when classifying regions that contain epithelial structures such as squamous epithelium.
Focusing on a single patch of a sample with false positive predictions, we used the Captum library~\citep{kokhlikyan2020captum} to apply popular post-hoc interpretability techniques.

\begin{figure*}[h!]
\floatconts
  {fig:lgc-avg-combined}
  {\caption{Averaged attribution heatmaps for two different pixels in the false positive region. The model consistently focuses on the nearby ground truth tumor area as an explanation for its incorrect prediction.}}
  {%
    \subfigure[Relative to pixel (150,150).]{\label{fig:lgc-averaged}%
      \includegraphics[width=0.35\textwidth]{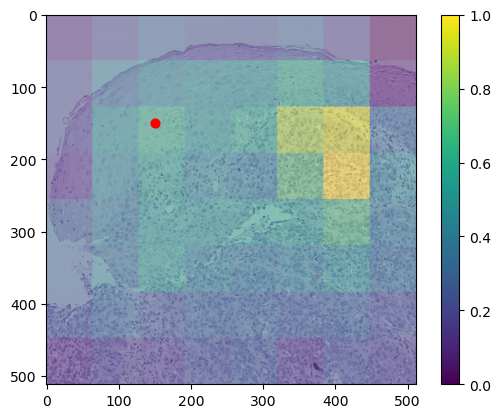}}%
    \qquad
    \subfigure[Relative to pixel (250,100).]{\label{fig:lgc-averaged-250}%
      \includegraphics[width=0.35\textwidth]{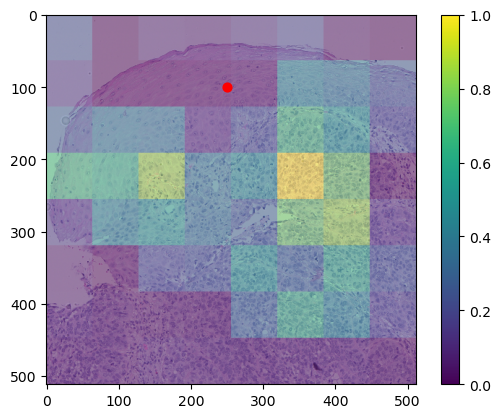}}
  }
\end{figure*}

Starting with Layer Grad-CAM~\citep{selvaraju2020grad}, we computed attributions for a pixel in a false positive region (e.g., at coordinates (150,150)) with respect to the final 2D convolutional layer in the encoder. 
The attributions were computed for each of the five models trained by nnU-Net and then averaged. 
While the individual attribution maps for each fold show noticeable disagreement (see \figureref{fig:lgc-5-fold}), hinting that the 5-fold split may be suboptimal for the available data, the averaged heatmap provides another explanation. 
As shown in \figureref{fig:lgc-averaged}, the model's confusion stems from its focus on the nearby tumorous region, suggesting that spatial proximity to the tumor is a primary cause for misclassifying adjacent epithelial tissue. 
This behaviour was consistent across different points in the false positive region.

\begin{figure*}[!h]
\floatconts
  {fig:lgc-5-fold}
  {\caption{Layer Grad-CAM attribution heatmaps for each of the five cross-validation models. Lightly coloured regions indicate positive contributions to the model's prediction that the red pixel at (150,150) is tumorous.}}
  {\includegraphics[width=\textwidth]{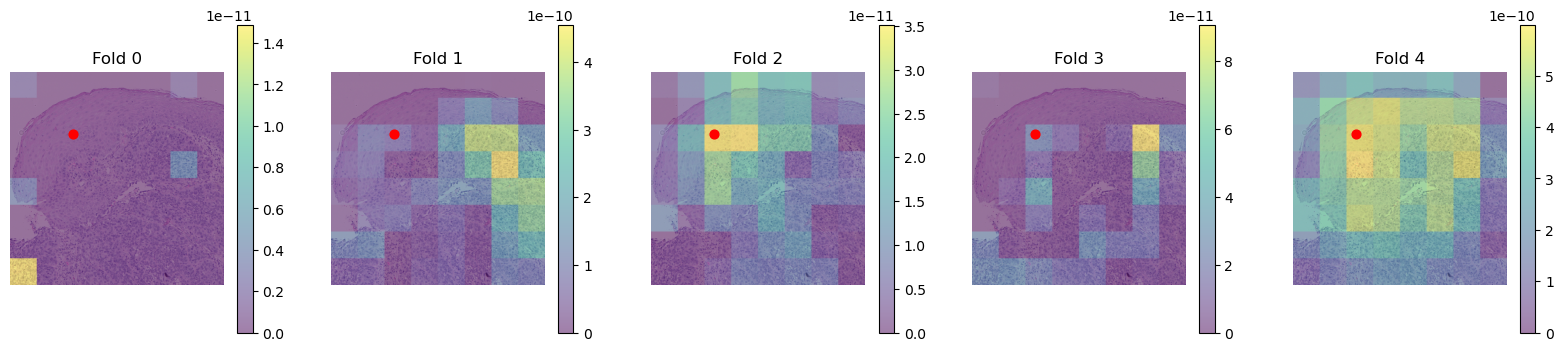}}
\end{figure*}

\begin{figure*}[htbp]
\floatconts
  {fig:ig-ablation}
  {\caption{Both Integrated Gradients (a) and Feature Ablation (b) point to channel 21 as a key contributor to the false positive prediction.}}
  {%
    \subfigure[Sum of absolute Integrated Gradient attributions for each channel on the ensemble model.]{\label{fig:ig-channel-attribution}%
      \includegraphics[width=0.45\textwidth]{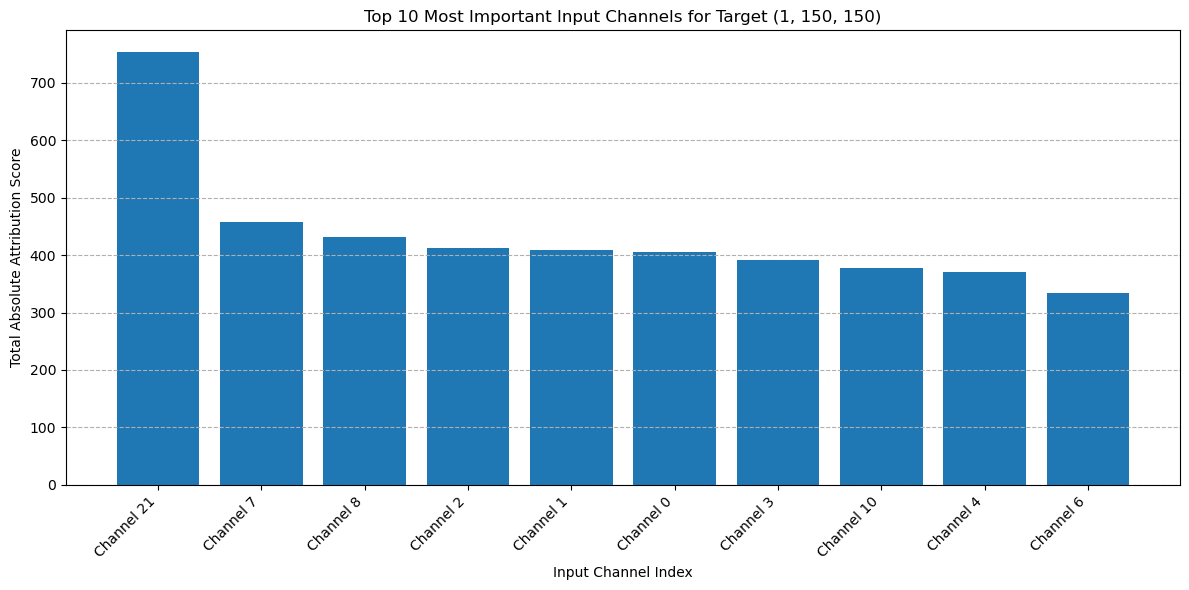}}%
    \qquad
    \subfigure[Mean attributions from Feature Ablation on the epithelium region.]{\label{fig:channel-ablation}%
      \includegraphics[width=0.45\textwidth]{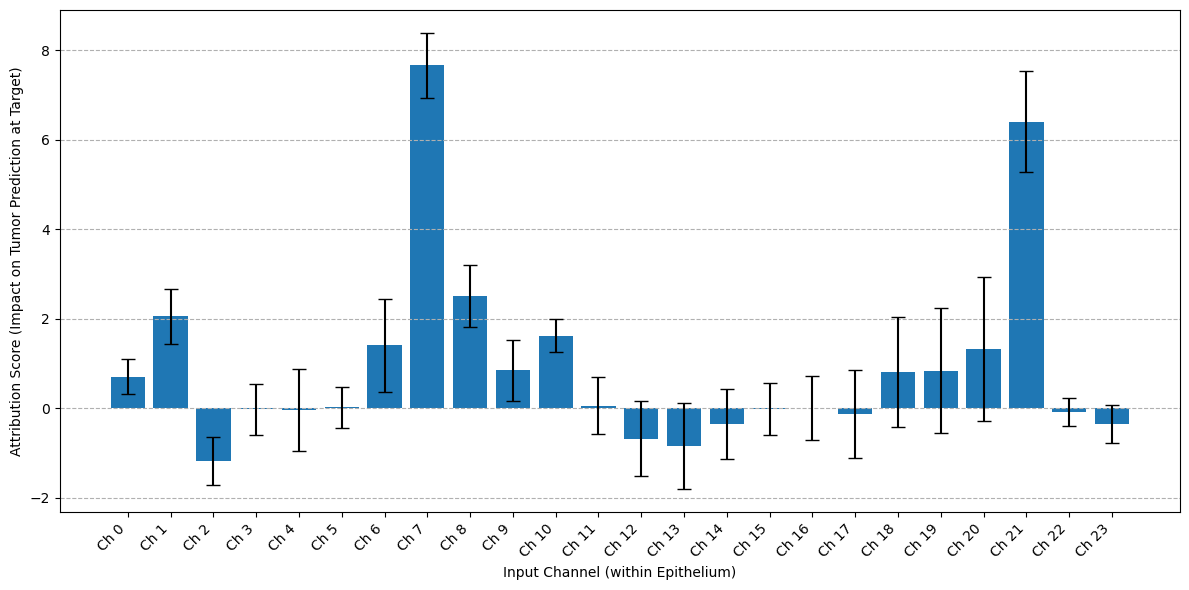}}
  }
\end{figure*}

To determine which input channels were most influential in this misclassification, we applied two further techniques. 
First, using Integrated Gradients~\citep{sundararajan2017axiomatic}, we computed the contribution of each of the 24 input channels to the prediction. By summing the absolute attribution values spatially for each channel, we obtained a contribution vector $c \in \mathbb{R}^{24}$. 
The results in \figureref{fig:ig-channel-attribution} highlight channel 21, a transmission channel intended for morphological context, as highly influential.
Secondly, we used Feature Ablation by masking the entire epithelial region for each channel and measuring the resulting change in the model's output for a target pixel. 
This confirmed the Integrated Gradients finding: \figureref{fig:channel-ablation} shows that channel 21, along with channel 7, provided significant positive evidence for the tumor class, contrary to the expectation that epithelial regions should provide negative evidence.

Visual inspection of the identified channels appears to corroborate this analysis, as we noticed the epithelial and tumorous regions are nearly indistinguishable, particularly in channel 21.
Squamous Cell Carcinoma is known for high morphological similarity between tumorous and healthy epithelial regions~\citep{mahadevan1996raman}. This suggests the model's confusion is rooted in the input data, where the spectral channels may be missing key morphological information needed to differentiate these similar tissue types.

\subsubsection{RamanSeg interpretability analysis}

The first step is to check if prototypes correspond to the classes they aim to represent.
Following the algorithm to identify the bounding region for a given prototype described by~\citep{chen2018}, we index into the corresponding segmentation mask, and compute the proportion of each class present in that section.
These proportions are visualized in \figureref{fig:class-proportions-ramanseg}.
Given that there are far more tumorous regions in the prototypes intended to represent tumour (the last fifteen prototypes), the prototypes learned by our training process must be meaningful.

\begin{figure}
\floatconts
  {fig:class-proportions-ramanseg}
  {\caption{Class proportions in prototype regions.}}
  {\includegraphics[width=0.8\linewidth]{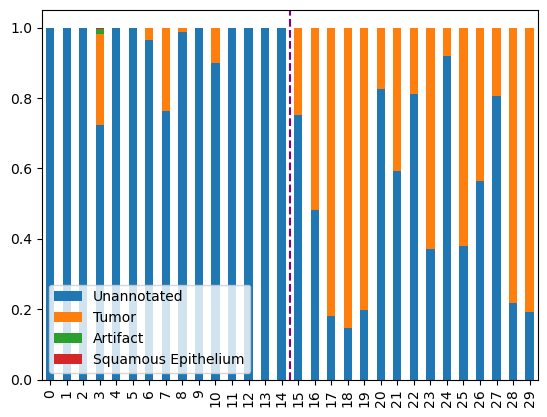}}
\end{figure}

Another property we verify is the diversity of the prototypes. The activation overlap loss we propose in \sectionref{sec:activation-overlap} should encourage prototypes to be diverse.
By plotting the \textit{inertia} of k-means clustering fits for increasing $k$, we can observe how useful additional clusters are in capturing the data.
\figureref{fig:kmeans-inertia} displays this result: given the steepness of the curve for the first fifteen fits, we can conclude that the majority of the prototypes are diverse, representing different regions. The activation overlap loss successfully encouraged prototype diversity.

\begin{figure}
\floatconts
  {fig:kmeans-inertia}
  {\caption{Inertia for each cluster fit $k$, along x-axis.}}
  {\includegraphics[width=0.7\linewidth]{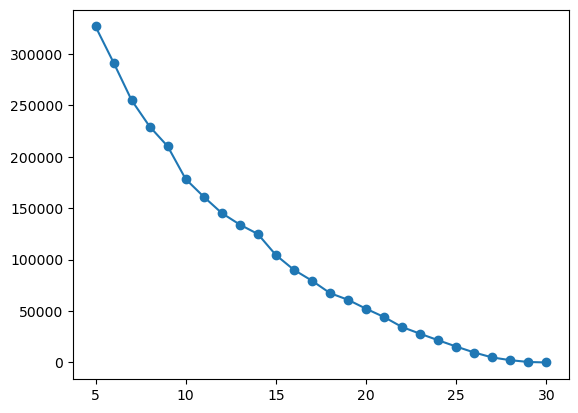}}
\end{figure}

Returning to the false positives triggered by healthy epithelium regions, while we could investigate individual similarity scores to understand why the model made such a prediction, there is a simpler method. Observing the types of prototypes learned from \figureref{fig:class-proportions-ramanseg}, it is clear that the training process has not included prototype(s) for squamous epithelium regions, which are important non-tumorous structures.

\section{Conclusion and future work}

This paper demonstrates the efficacy of using multi-channel Raman spectroscopy data for the semantic segmentation of cancerous tissue.
Our primary model, an nnU-Net ensemble, achieved a Dice score of 80.9\% on the holdout set, setting the state-of-the-art performance on the hyperspectral dataset. 
We investigated the model's primary failure mode --- confusion between tumorous and epithelial tissue --- using post-hoc interpretability methods. 
Integrated Gradients revealed this confusion was largely driven by misleading morphological similarities in the transmission channel, highlighting the utility of such techniques for model debugging.
To create a more inherently interpretable model, we proposed RamanSeg, a novel prototype-based architecture. A projection-free variant outperformed a baseline U-Net with a Dice score of 67.3\%.
RamanSeg supported a straightforward way to understand the false-positive failure mode: checking the class proportions of the regions of the learned prototypes.

\acks{This work was supported by EU HORIZON EUROPE research projects TROPHY (101047137) and CHARM (101058004). It was also supported by the German Cancer Aid project ARBOR (Deutsche Krebshilfe Grant No. 70116061, 70116281) and Federal Ministry of Research, Technology and Space project HyperTUM (BMFTR No. 16SV9564).}

\bibliography{paper}


\end{document}